\documentclass[dvips]{acta}
\usepackage{supertabular,lscape,epsfig}
\usepackage{amssymb}
\usepackage{amsmath}
\SetPages{1}{12}
\SetVol{59}{2009}

\begin{document}

\begin{Titlepage}

\Title { On the Origin of Tilted Disks and Negative Superhumps }

\Author {J.~~S m a k}
{N. Copernicus Astronomical Center, Polish Academy of Sciences,\\
Bartycka 18, 00-716 Warsaw, Poland\\
e-mail: jis@camk.edu.pl }

\Received{  }

\end{Titlepage}

\Abstract { 
The origin of tilted disks in cataclysmic variables is explained in terms 
of a model involving the stream-disk interactions. 
Tilted, precessing disk causes periodically variable asymmetry in the 
irradiation of the two hemispheres of the secondary component, 
resulting in variable vertical component of the velocity of the stream. 
The following stream-disk interactions provide additional vertical acceleration 
to disk elements needed to produce and maintain disk tilt. 
Predictions based on this model compare favorably with observations. 
} 
{\it accretion, accretion disks -- binaries: cataclysmic variables,
stars: dwarf novae }

\section { Introduction }

Negative superhumps are quite common among cataclysmic variables. 
They are present in dwarf novae of the SU UMa type during their superoutbursts 
and also among systems with stationary accretion -- the permanent 
{\it negative} superhumpers (see Patterson 1999, Wood and Burke 2007, 
Olech et al. 2009 and references therein). 
Their periods are shorter than the orbital periods, the corresponding 
"period deficit" $\epsilon$ being correlated with the orbital period. 
There are examples of negative superhumpers showing also the {\it common} 
superhumps, the two types either being present simultaneously or switching 
from one type to another. In those cases the "period excess" of {\it common} superhumps and the "period deficit" of {\it negative} superhumps are 
correlated (cf. Olech et al. 2009).   

The commonly accepted interpretation of negative superhumps explains them 
in terms of a tilted, precessing disk (cf. Patterson et al. 1993, 
Harvey et al. 1995, Patterson 1999, and references therein). 
Supporting this interpretation are: (1) the presence of light variations 
with $P_{prec}$ observed commonly in such systems and (2) the fact that 
precession periods obtained from the observed values of $P_{nsh}$ and 
$P_{orb}$ agree reasonably well with theoretical predictions 
(see Larwood et al. 1996 and references therein). 

Patterson et al. (1997) were the first to suggest that negative superhumps 
could be due to the "spot" produced by the variable stream impact as 
it transits across the surface of a tilted disk.   
More recently this was confirmed by Wood and Burke (2007) who used the 3D SPH simulations of a tilted, precessing disk to produce light curves closely 
resembling the observed superhumps (see also Montgomery 2009 and Wood et al. 2009). 

The origin of the disk tilt, however, continues to be unknown. 
In particular, models proposed for other, more exotic objects (such as HZ Her 
or SS 433), are not applicable to the case of disks in cataclysmic variables 
(cf. Wood and Burke 2007 and references therein). 

In the present paper we propose to explain the origin of tilted disks 
in cataclysmic variables in terms of a model involving the stream-disk 
interactions in a situation when the tilted, precessing disk causes 
variable irradiation of the two hemispheres of the secondary component and 
the resulting stream has periodically variable vertical velocity component.  

We begin, in Section 2, with definitions and formulae to be used 
in further sections. Section 3 describes the irradiation controlled mass 
outflow and the next two sections are devoted to a detailed discussion of 
the resulting stream: its trajectory (Section 4) and its collision with 
the surface and edge of the disk (Section 5). 
The model, involving the stream-disk interactions, is presented in Section 6, 
its predictions being compared with observations in Section 7.

\section {Definitions and Formulae }

The precession period of a tilted disk is related to the orbital period 
and the negative superhump period by

\beq
{1\over {P_{prec}}}~=~{1\over {P_{nsh}}}~ - ~{1\over {P_{orb}}}~. 
\eeq 

\noindent
Accordingly we have 

\beq
\phi_{prec}~=~\phi_{nsh}~-~\phi_{orb}~,  
\eeq 

\noindent 
with the zero-points of phases being defined as follows: 
$\phi_{orb}=0$ at conjuction (in particular -- at eclipse) and 
$\phi_{nsh}=0$ at superhump maximum.  

It will also be useful to recall that 

\beq
{{d\phi_{nsh}}\over {d\phi_{orb}}}~=~{1\over {1-\epsilon}}~, 
\eeq

\noindent
where

\beq
\epsilon~=~{{P_{orb}~-P_{nsh}}\over P_{orb}}~, 
\eeq

\noindent
is the negative superhump period deficit. 

Let us now consider a disk with radius $r_d$, geometrical thickness $z/r$, 
and tilt angle $\delta$. Its geometry (see Fig.1) is described by the following 
set of equations: 

\beq
z_d~=~-~z_\circ~\cos \theta~, 
\eeq

\noindent
where $\theta$ is the position angle on the surface of the disk and 

\beq
z_\circ~=~r_d~\sin~\delta~, 
\eeq

\noindent
and 

\beq
z_t~=~z_d~+~\Delta z_d~, ~~~~~ z_b~=~z_d~-~\Delta z_d~, 
\eeq 

\beq
\Delta z_d~=~r_d~(z/r)~.  
\eeq 

\begin{figure}[htb]
\epsfysize=1.5cm 
\hspace{2.0cm}
\epsfbox{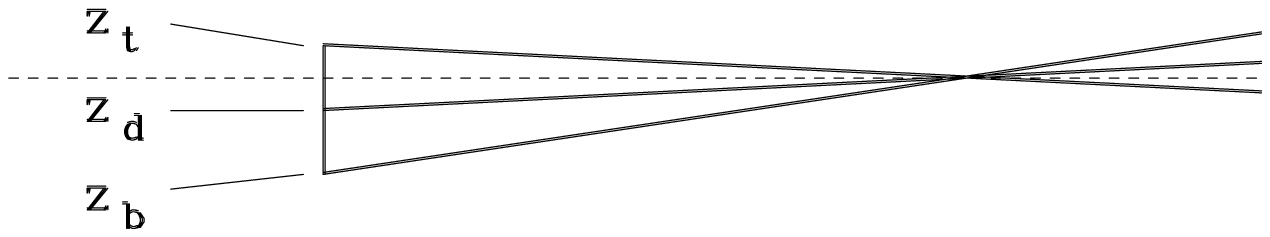} 
\vskip 5truemm
\FigCap { Geometry of the tilted disk with $\delta=3^\circ$ (see Section 7.1) 
and $z/r=0.10$ at $\theta=0$. Dotted line marks the orbital plane.  }
\end{figure}

Two comments are worth adding here. First, that the specific value of the 
disk radius $r_d$ is actually irrelevant since in what follows we will be 
dealing only with {\it angles}. 
Secondly, that all angles will be expressed here in units of phase, so that  
$\theta=1$ corresponds to $2\pi$ or $360^\circ$. 

Let us now consider an element of the outer disk. Its motion in the z-coordinate 
is described by Eq.(5) with 

\beq
\theta~=~{{2 \pi}\over {P_d} } ~t~, 
\eeq

\noindent
where $P_d$ is the period of oscillations (equal also to the period of revolution). 

The acceleration, calculated from Eqs.(5) and (9), is    

\beq
{{d^2 z}\over {dt^2}}~=~z_\circ~\left({{2\pi}\over{P_d}}\right)^2
\cos \theta~+~2~{{dz_\circ}\over {dt}} 
\left({{2\pi}\over{P_d}}\right) \sin \theta~. 
\eeq

The second term on the right hand side of this equation implies that to produce 
and maintain disk tilt we need a mechanism capable of providing additional 
acceleration of the form  

\beq
a_z~\sim~\sin \theta~. 
\eeq

\section { The Irradiation Controlled Mass Outflow }

It has recently been shown (Smak 2008) that superoutbursts of Z Cha (and most 
likely of other dwarf novae) are due to a major enhancement in the mass outflow 
rate. This provides a strong argument in favor of the concept of irradiation 
controlled mass outflow. The details of this mechanism, however, still remain 
controversial (see Smak 2009b). 
The problem here is connected with the fact that the equatorial parts of the 
secondary, including L$_1$, are in the shadow cast by the disk. 
The crucial question then is whether the material from irradiated regions 
flowing towards L$_1$ is still hot enough after reaching this point to produce 
substantial modulation of the mass outflow rate. 
Assuming that this is indeed the case it was possible to propose a new 
interpretation for superhumps (Smak 2009b). 
The simplicity and self-consistency of this interpretation could, in fact, 
be treated as an indirect argument in favor of this assumption. 

In what follows we also assume that the outflow from L$_1$ is controlled 
by irradiation and, in particular, that in the case of variable irradiation 
there is a time delay between irradiation and the resulting dissipation 
at the point of impact (see Smak 2009b): 

\beq
\Delta t~=~\Delta t_{flow}~+~\Delta t_{str}~, 
\eeq

\noindent
where $\Delta t_{flow}$ is the time needed for the flow from irradiated regions 
to reach L$_1$, and $\Delta t_{str}$ is the time needed for the stream to reach 
the point of impact. 
The flow time $\Delta t_{flow}$ depends, in general, on several parameters; 
in particular it depends on the distance between L$_1$ and the shadow boundary.

\section { Stream Trajectories }

When the disk is coplanar with the orbital plane the initial velocity vector 
of the stream at L$_1$ is of the form $\vec v_\circ=[v_{x,\circ},v_{y,\circ},0]$. 
Let us, however, consider the case of a tilted disk when its part facing 
the secondary is tilted below the orbital plane (in the notation used 
in Section 6 this corresponds to $\theta_{2,d}=0$). 
Compared to the coplanar case, the shadow boundary on the top hemisphere 
of the secondary is now closer to L$_1$, while that on the bottom hemisphere -- 
further away from L$_1$. As a result, the contribution to the mass outflow from 
the material flowing from the top hemisphere is dominant and, consequently, 
the initial velocity vector now has a non-zero vertical component: $v_{z,\circ}<0$. 

To study the consequences of such a situation we calculate stream trajectories in 
three dimensions. For the initial stream velocity at L$_1$ we adopt various combinations of its components (in dimensionless units): 
$v_{x,\circ}=0.01-0.05$, $v_{y,\circ}=(-0.01)-(-0.05)$, and $v_{z,\circ}=(-0.02)-(-0.05)$. 
Shown in Fig.2 are results obtained for the mass ratio $q=0.3$ which corresponds 
to the typical periods of negative superhumpers, as for example those listed 
below in Table 1. (Results obtained with other mass ratios were qualitatively 
the same).

\begin{figure}[htb]
\epsfysize=10.0cm 
\hspace{2.5cm}
\epsfbox{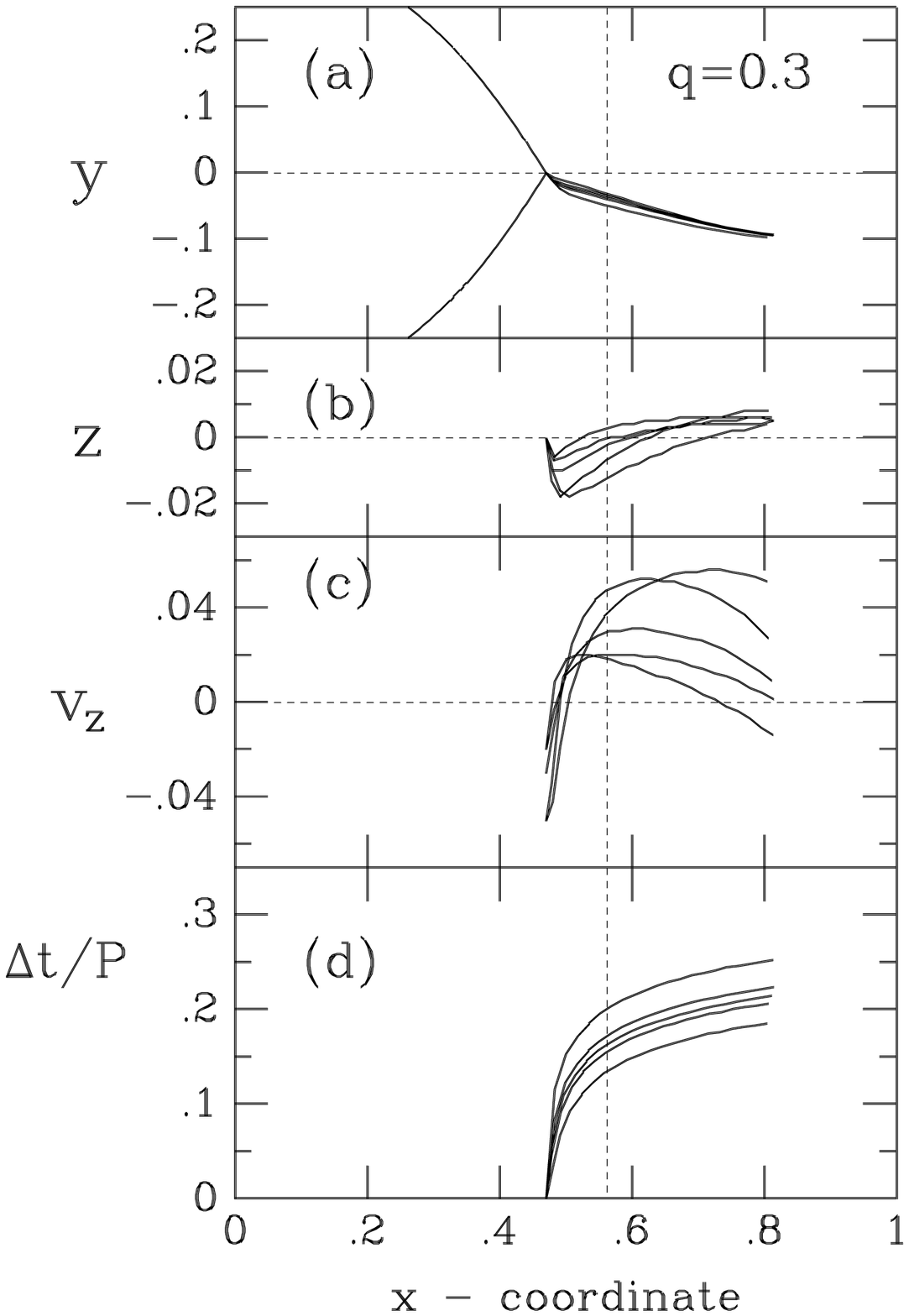} 
\vskip 5truemm
\FigCap { ({\it a}) Stream trajectories in projection on the x-y plane. 
({\it b}) Stream trajectories in projection on the x-z plane.  
({\it c}) Vertical velocity component of the stream. 
({\it d}) The "flight" time $\Delta t_{str}$ in units of the orbital period. 
Vertical dotted line in all plots marks the location of the disk edge. }
\end{figure}

There are three important conclusions to be drawn from Fig.2: 
(1) The stream remains very close to the orbital plane. 
(2) The "flight" time from L$_1$ to $r=r_d$ is $\Delta t_{str}\approx 0.15 P_{orb}$ 
and depends only weakly on the initial conditions. 
(3) At the point of impact the sign of the vertical velocity component of the 
stream is {\it opposite} to that of the initial velocity.

\section { The Stream Overflow }

The stream overflow was originally expected (cf. Smak 1985, Hessman 1999) 
to occur when the disk is geometrically thin, i.e. mainly in quiescent dwarf novae. 
Evidence is now available, however, for a substantial stream overflow in Z Cha 
during its superoutbursts (Smak 2007, 2009a). 
Besides, there is also theoretical evidence (Kunze et al. 2001) suggesting 
that this is indeed a much more common phenomenon. 

Using disk geometry described in Section 2 we can calculate the fraction $f_t$ 
of the stream material overflowing the top part of the disk and the fraction $f_e$ 
colliding with disk edge. 
We adopt: $z/r=0.10$ (cf. Smak 1992), $\delta=3^\circ$ (see Section 7.1) 
and the density distribution in the stream given by the two-dimensional Gaussian 
formula with $\sigma/\Delta z_d=0.5$, 1.0, and 2.0. With these assumptions we have 

\beq
f_{t,e}~=~\int _{z_1}^{z_2} {1 \over {2 \pi \sigma}}~{\rm exp}~
\left( -z^2/2\sigma^2 \right)~dz~, 
\eeq 

\noindent
where the integration limits are: $z_1=z_t$ and $z_2=\infty$ for $f_t$, 
and $z_1=z_b$ and $z_2=z_t$ for $f_e$.

\begin{figure}[htb]
\epsfysize=4.5cm 
\hspace{1.5cm}
\epsfbox{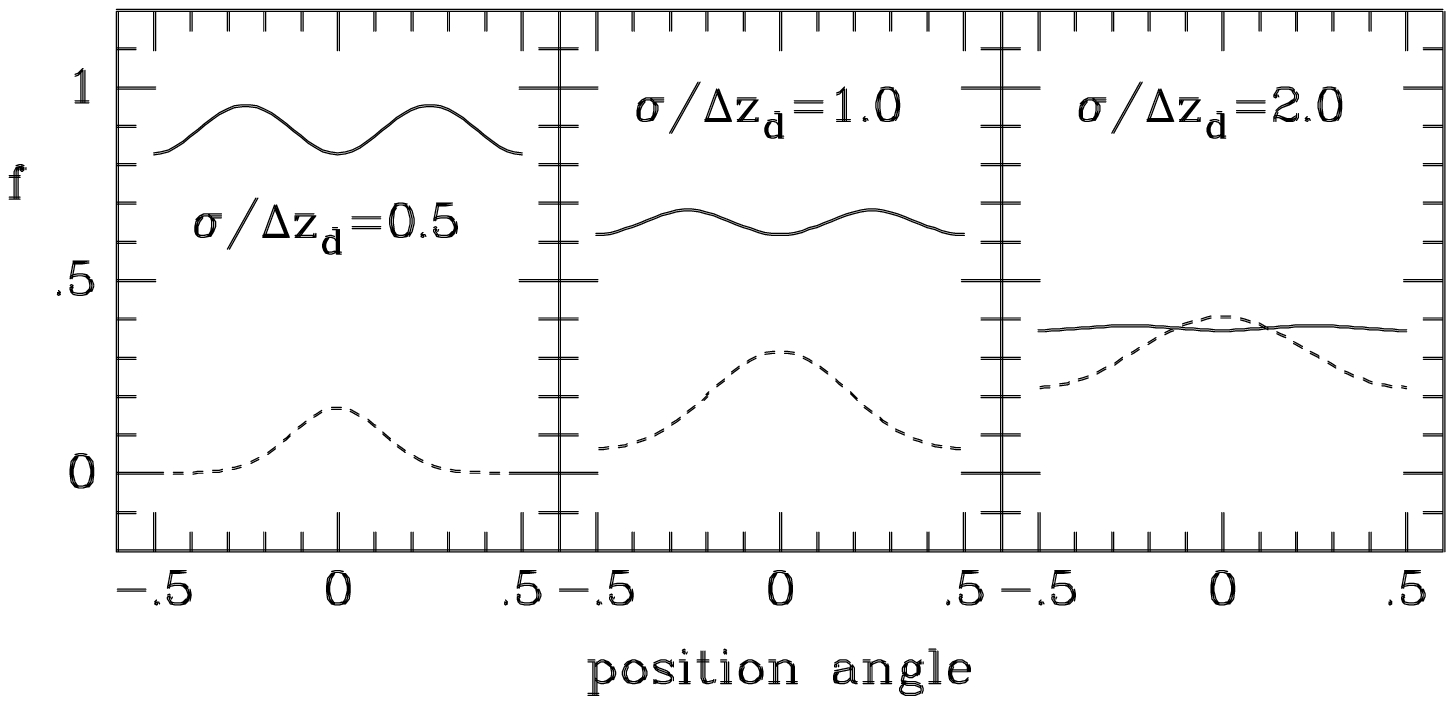} 
\vskip 5truemm
\FigCap { The fraction $f_t$ of the stream material overflowing the 
top part of the disk (broken lines) and the fraction $f_e$ colliding with 
its edge (solid lines) are shown as a function of the position angle  
for $\delta=3^\circ$ and for three values of $\sigma$. }
\end{figure}

Results are presented in Fig.3. As could be expected, 
substantial overflow occurs when $\sigma$ is comparable to $\Delta z_d$. 
In particular, $f_t$ becomes larger than $f_e$ for $\sigma>2\Delta z_d$. 
Furthermore, variations of $f_t$ show only one maximum (and one minimum) per 
cycle, while those of $f_e$ -- two maxima (and two minima) per cycle 
(we shall return to this in Section 7.2). 

Additional calculations were also made to reproduce the simple situation 
considered by Wood and Burke (2007) with the stream fully overflowing one side 
of the disk ($f_t\approx 1$). We find that this would require either much 
larger values of $\sigma$ (which is rather unlikely) or tilt angles much larger 
than $\delta\sim 6^\circ$ (in fact, Wood and Burke used $\delta=5^\circ$, but 
their disk was much thinner). We shall return to such a situation in Section 7.1.

\section { The Model }

The stream, colliding with the disk with non-zero vertical velocity component, 
is an obvious source of extra acceleration of disk elements. 
We propose that, under suitable conditions, this can be the mechanism capable 
of producing and maintaining disk tilt. 

The geometry of irradiation depends on the relative orientation of the 
tilted disk with respect to the secondary component. 
The position angle of the secondary with respect to the observer is 

\beq
\theta_{2,obs}~=~\phi_{orb}^*~. 
\eeq

\noindent
The asterisk $*$ is used here (and below) to emphasize that the phase refers  
to the moment of irradiation of the secondary component. 
The position angle of the lowest point of the disk, corresponding to $\theta=0$ 
(as defined above), with respect to the observer is 

\beq
\theta_{d,obs}~=~\phi_{prec}^*~-~\theta_\circ~, 
\eeq

\noindent
where $\theta_\circ$ is the precession phase at which the point with 
$\theta=0$ is facing the observer. 
Note that $\theta_{d,obs}$, which changes due to retrograde precession,  
is counted in the direction opposite to $\theta_{2,obs}$. 
With this definition, the position angle of the disk point with $\theta=0$ 
with respect to secondary component is 

\beq
\theta_{d,2}~=~\theta_{d,obs}~+~\theta_{2,obs}~
=~(\phi_{prec}^*~+~\phi_{orb}^*)~-~\theta_\circ~
=~\phi_{nsh}^*~-~\theta_\circ~.  
\eeq

Consequently, the vertical component of the terminal stream velocity at 
the point of impact $v_{imp}$ can be expressed as a function of $\phi_{nsh}^*$.
Assuming, for simplicity, that its variations are cosinusoidal we can write   

\beq
v_{imp}~=~v_{imp,\circ}~\cos ~(~\phi_{nsh}^*~-~\theta_\circ~)~, 
\eeq

\noindent
where $v_{imp,\circ}>0$ and corresponds to $\phi_{nsh}^*=\theta_\circ$ 
(see Fig.2c and Eq.16). The additional acceleration produced by the stream 
is then 

\beq
a_{z,str}~=~{\dot M}~v_{imp,\circ}~\cos ~(~\phi_{nsh}^*~-~\theta_\circ~)~
          \sim~\sin ~[~(~\phi_{nsh}^*~-~\theta_\circ~)~+~0.25~]~.  
\eeq

The moment of impact is delayed with respect to the moment of irradiation 
by $\Delta t$ (see Eq.12), the corresponding phase delays being  
$\Delta\phi_{orb}=\Delta t/P_{orb}$ (see Fig.2d) and 
$\Delta\phi_{nsh}=\Delta\phi_{orb}/(1-\epsilon)$ (see Eq.3). 
The phase of impact is then 

\beq
\phi_{nsh}~=~\phi_{nsh}^*~+~\Delta \phi_{nsh}~
=~\phi_{nsh}^*~+~{{\Delta \phi_{orb}}\over {1-\epsilon}}~, 
\eeq

\noindent
and the position angle on the disk at the point of impact 

\beq
\theta~=~\phi_{nsh}^*~-~\theta_\circ~+~\Delta \phi_{nsh}~+~\alpha~ 
=~\phi_{nsh}^*~-~\theta_\circ~
+~{{\Delta \phi_{orb}}\over {1-\epsilon}}~+~\alpha~, 
\eeq
   
\noindent
where $\alpha$ is the position angle of the impact point with respect to 
the line joining the two components. For $q=0.3$ and at $r=r_d$ we have 
$\alpha\approx 0.014$ (in phase units).  

The acceleration needed to produce and maintain disk tilt (Eq.11 in 
Section 2) must then be 

\beq
a_z~\sim ~\sin \left [~(~\phi_{nsh}^*~-~\theta_\circ~)~
+~{{\Delta \phi_{orb}}\over {1-\epsilon}}~+~\alpha ~\right ]~. 
\eeq

Comparing Eqs.(18) and (21) we immediately conclude that our proposed 
mechanism is most efficient when 

\beq
{{\Delta \phi_{orb}}\over {1-\epsilon}}~+~\alpha~=~0.25~,  
\eeq

\noindent
or, using $\epsilon\approx 0.02$ and $\alpha\approx 0.014$, when  

\beq
\Delta \phi_{orb}\approx 0.23~.
\eeq

In general the efficiency of this mechanism is described by the sign and value 
of the integral

\bdm
W~= \int _0^1 \cos ~(\phi_{nsh}^*~-~\theta_\circ)~
\sin \left[ ( \phi_{nsh}^*~-~\theta_\circ~) ~+~
{{\Delta \phi_{orb}}\over {1-\epsilon}}~+~\alpha \right]~ 
{\rm d}~(\phi_{nsh}^*~-~\theta_\circ)~=
\edm

\vskip -5truemm

\beq
=~\pi~\sin \left( {{\Delta \phi_{orb}}\over {1-\epsilon}}+\alpha \right)~,   
\eeq

\noindent
which implies that the mechanism is effective ($W>0$) for  

\beq
-0.02~<~\Delta \phi_{orb}~<~0.48~.  
\eeq

The value of $\Delta \phi_{orb}$, estimated earlier (Smak 2009b) for 
three systems (IY UMa, DV UMa and OY Car) showing {\it common} superhumps,  
was found to be $\Delta \phi_{orb}=\Delta t/P_{orb}\approx 0.66$  
which is clearly outside the range required by Eq.(25). This is not surprising, 
however, since those three objects do {\it not} show negative superhumps. 

In the case of {\it negative} superhumpers with tilted disks 
the shadow boundary on the top (or bottom) hemisphere of the secondary 
is much closer to L$_1$ and this makes the flow time (Eq.12) much shorter. 
Consequently the value of $\Delta \phi_{orb}$ can easily be small enough 
to fulfill condition imposed by Eq.(25). 
To illustrate this point let us consider the extreme case, when the shadow 
boundary is close to L$_1$ (see Section 7.1). 
In such a case $\Delta t_{flow} \rightarrow 0$ and $\Delta \phi_{orb}$ becomes 
as small as $\Delta t_{str}/P_{orb}\approx 0.15$ (see Eq.12 and Fig.2d).

\section { Model Predictions } 

\subsection { Disk Tilts } 

Let us consider the situation when the tilt is so large that the vicinity 
of L$_1$ is fully exposed to direct irradiation. 
In such a case the vertical component of the stream velocity becomes 
$v_{z,\circ}\equiv 0$ and the stream mechanism discussed in the previous 
Section no longer operates. 
This means that there is a natural upper limit to the tilt angle given by $\delta_{max}={\rm arc} \sin (z/r)$. Using values of $z/r=0.10-0.12$ typical 
for hot disks we predict $\delta_{max}\approx 6-7^\circ$. 

To test this prediction we analyze the best documented examples of negative 
superhumpers which show periodic light variations with $P_{prec}$. 
They are listed in Table 1, where the second and third columns contain the semi-amplitudes $A$ of those variations and the orbital inclinations;  
in the case of non-eclipsing systems, with no estimates of inclinations 
being available, we adopt a range: $i=20-60^\circ$.

\begin{table}[h!]
{\parskip=0truept
\baselineskip=0pt {
\medskip
\centerline{Table 1}
\medskip
\centerline{ Semi-Amplitudes and Tilt Angles }
\medskip
$$\offinterlineskip \tabskip=0pt
\vbox {\halign {\strut
\vrule width 0.5truemm #&	
\quad\hfil#\quad&	        
\vrule#&			
\quad\hfil#\hfil\quad&          
\vrule#&			
\quad\hfil#\hfil\quad&	        
\vrule#&			
\quad#\hfil\quad&	        
\vrule#&			
\quad\hfil#\hfil\quad&	        
\vrule width 0.5 truemm # \cr	
\noalign {\hrule height 0.5truemm}
&&&&&&&&&&\cr
&Star\hfil&& $A{\rm (mag)}$ && $i{\rm(deg)}$ && Refs. && $\delta{\rm(deg)}$ &\cr
&&&&&&&&&&\cr
\noalign {\hrule height 0.5truemm}
&&&&&&&&&&\cr
&    PX And && 0.20 && 74  && 1,2     && 3 &\cr
&  V603 Aql && 0.07 && 20  && 3       && 7 &\cr
&    TT Ari && 0.07 && 29  && 4,5,6,7 && 4 &\cr
&    TV Col && 0.20 && 70  && 8,9,10  && 3 &\cr
&  V751 Cyg && 0.05 &&20-60&& 11      &&5-1&\cr
& V1084 Her && 0.08 &&20-60&& 12      &&6-2&\cr
&  V442 Oph && 0.10 && 67  && 12,13   && 2 &\cr
&&&&&&&&&&\cr
\noalign {\hrule height 0.5truemm}
}}$$
References: 
1. Stanishev et al. (2002, Fig.5), 2. Thorstensen et al. (1991), 
3. Patterson et al. (1997, Table 1), 4. Semeniuk et al. (1987, Fig.9), 
5. Kraicheva et al. (1997), 6. Kraicheva et al. (1999), 
7. Wu et al. (2002), 8. Barrett et al. (1988, Fig.4), 9. Hellier (1993),
10. Retter et al. (2003, Table 4), 11. Patterson et al. (2001, Fig.3), 
12. Patterson et al. (2002, Figs.8 and 10), 13. Ritter and Kolb (1998). 
}}
\end{table}

The semi-amplitude $A$ (in magnitudes) can be written as  

\beq
A~=~{{dM_V}\over {di}}~\delta~,
\eeq

\noindent
where $dM_V/di$ describes the dependence of the observed luminosity of the disk 
on its inclination (or -- generally -- on the viewing angle). 

At inclinations $i\leq75^\circ$ it is sufficient to use the flat disk 
approximation giving 

\beq
L_d(i)~=~<L_d>~{6\over {3-u}}~(~1~-~u~+~u~\cos i~)~\cos i~,
\eeq 

\noindent
where $u$ is the limb darkening coefficient for which we adopt $u=0.6$. 
Turning to magnitudes we get 

\beq
{{dM_V}\over {di}}~=~0.01895~
{{(~1~-~u~+~2~u~\cos i~)~\sin i}\over
{(~1~-~u~+~u~\cos i~)~\cos i}}~[{\rm mag/deg}]~. 
\eeq 

The resulting values of $\delta$ are listed in the last column of Table 1. 
As one can see they are smaller that $\delta_{max}$ predicted above, 
the typical tilt being $\delta\approx 3^\circ$.

\subsection { The Superhump Light Curves } 

The negative superhumps are thought to be due -- primarily -- to the collision 
of the stream with the surface of the tilted, precessing disk. 
There are two other effects, however, which must be considered. 
First, that part of the stream collides with disk edge (see Section 5). 
Secondly, that the outflow rate $\dot M$ is likely to vary with variable 
geometry of irradiation, i.e. with $\theta_{d,2}$, or with $\phi_{nsh}^*$. 

Therefore the shape of the light curve can be formally written as 

\beq
\ell(\phi_{nsh})~\sim~\dot M(\phi_{nsh}^*)~[~f_e(\phi_{nsh})~+~x~f_t(\phi_{nsh})~]~,
\eeq

\noindent
where $\phi_{nsh}^*$ is the phase at the moment of irradiation, 
$\phi_{nsh}$ -- the phase at the moment of impact, and 
"x" in front of $f_t$ is intended to represent the higher value of the 
impact parameter $\Delta V^2$ in the case of the overflowing parts of the stream. 

Limiting our discussion to qualitative considerations we recall (see Fig.3 
in Section 5) that variations of $f_t$ show one maximum (and one minimum) 
per cycle while those of $f_e$ -- two maxima (and two minima) per cycle. 
Variations of $\dot M$ are also expected to show two maxima (and two minima) 
per cycle. Taking this into account we conclude that 
(1) the contribution from overflowing parts of the stream $f_t$ 
must be dominant (due obviously to $x> 1$; see above), and 
(2) the contributions from variations of $f_e$ and $\dot M$ make the shape 
of the light curve significantly different from a simple cosine wave. 

The second prediction is confirmed (at least qualitatively) by the observed 
shapes of negative superhump light curves which in nearly all cases 
(Barett et al. 1988, Fig.4; Patterson 2001, Fig.3; Patterson et al. 1997, Fig.4; 
Patterson et al. 2002, Figs.8,10,15; Stanishev et al. 2002, Fig.3) 
are clearly non-cosinusoidal.

\subsection { The Light Variations with $P_{prec}$ } 

The negative superhump maximum, which occurs -- by definition -- at 
$\phi_{nsh}=0$, is produced when the stream is overflowing the top part of 
the disk at the position angle $\theta=0$. Using Eqs.(19) and (20) we get  

\beq
\theta_\circ~=~\alpha~. 
\eeq 

As mentioned in the Introduction, the mean luminosity of the disk varies 
with $P_{prec}$. The maximum of those variations occurs when $\theta_{d,obs}=0$. 
Combining Eqs.(15) and (30) we then predict that it should ocur at  

\beq
\phi_{prec}~=~\alpha~\approx~0.014~.  
\eeq 

\noindent
This agrees nicely with observations (e.g. TT Ari -- Semeniuk et  al. 1987, 
TV Col -- Hellier 1993 and PX And -- Stanishev et al. 2002) which show that 
the maximum occurs at $\phi_{prec}\approx 0$.

\section { Discussion }

The model presented in this paper appears quite simple and self-consistent. 
Its predictions compare favorably with observations. 
There are several problems and questions, however, which should be answered 
prior to considering it as fully acceptable. 

(1) What makes the originally coplanar disk to become tilted? 
The qualitative answer to this question may be quite simple: 
when some part of the outer disk -- due to a random fluctuation -- 
is deflected above or below the orbital plane, then the mechanism described 
above can begin to operate. What conditions, however, are 
necessary for producing the positive feed-back (especially for condition 
given by Eq.25 to be fulfilled)? 

(2) What is the cause of transitions from a tilted disk (with negative 
superhumps) to a coplanar disk (with common superhumps) -- 
and {\it vice-versa} -- observed in many systems (e.g. in V603 Aql or TT Ari; 
see references to Table 1)? 

(3) How can we explain the simultaneous presence of negative and common 
superhumps in some systems (e.g. in V603 Aql or V503 Cyg; see references to 
Table 1)? And why are such cases so rare?

\begin {references} 

\refitem {Barrett, P., O'Donoghue, D., Warner, B.} {1988} {\MNRAS} {233} {759}

\refitem {Harvey, D., Skillman, D.R., Patterson, J., Ringwald, F.A.} {1995} 
         {\PASP} {107} {551}

\refitem {Hellier, C.} {1993} {\MNRAS} {264} {132} 

\refitem {Hessman, F.V.} {1999} {\ApJ} {510} {867} 

\refitem {Kraicheva, Z., Stanishev, V., Iliev, L., Antov, A., Genkov, V.}
         {1997} {\AAS} {122} {123}

\refitem {Kraicheva, Z., Stanishev, V., Genkov, V., Iliev, L.}
         {1999} {\AA} {351} {607}

\refitem {Kunze, S., Speith, R., Hessman, F.V.} {2001} {\MNRAS} {322} {499} 

\refitem {Larwood, J.D., Nelson, R.P., Papaloizou, J.C.B., Terquem, C.} {1996}
         {\MNRAS} {282} {597}

\refitem {Montgomery, M.M.} {2009} {\MNRAS} {394} {1897}  

\refitem {Olech, A., Rutkowski, A., Schwarzenberg-Czerny, A.} {2009} 
         {\MNRAS} {~} {{\it in press}} 

\refitem {Patterson, J.} {1999} {{\it Disk Instabilities in Close Binary Systems}, 
         Eds. S.Mineshige and J.C.Wheeler (Tokyo: Universal Academy Press)} {~} {61}

\refitem {Patterson, J., Thomas, G., Sklillman, D.R., Diaz, M.} {1993} 
         {\ApJS} {86} {235}

\refitem {Patterson, J., Kemp., J., Saad, J., Skilmann, D.R., Harvey, D., Fried, R.,
          Thorstensen, J.R., Ashley, R.} {1997} {\PASP} {109} {468}

\refitem {Patterson, J., Thorstensen, J.R., Fried, R., Skillman, D.R., Cook, L.M.,
          Jensen, L.} {2001} {\PASP} {113} {72}

\refitem {Patterson, J. et al.} {2002} {\PASP} {114} {1364} 

\refitem {Retter, A., Hellier, C., Augusteijn, T., Naylor, T., Bedding, T.R., 
          Bembrick, C., McCormick, J., Velthuis, F.} {2003} {\MNRAS} {340} {679}

\refitem {Ritter, H., Kolb, U.} {1998} {\AAS} {129} {83}

\refitem {Semeniuk, I., Schwarzenberg-Czerny, A., Duerbeck, H., Hoffmann, M., 
          Smak, J., St{\c e}pie{\'n}, K., Tremko, J.} {1987} {\Acta} {37} {197}

\refitem {Smak,J.} {1985} {\Acta} {35} {351}

\refitem {Smak,J.} {1992} {\Acta} {42} {323}

\refitem {Smak,J.} {2007} {\Acta} {57} {87}

\refitem {Smak,J.} {2008} {\Acta} {58} {55} 

\refitem {Smak,J.} {2009a} {\Acta} {59} {109} 

\refitem {Smak,J.} {2009b} {\Acta} {59} {121} 

\refitem {Stanishev, V., Kraicheva, Z., Boffin, H.M.J., Genkov, V.} 
         {2002} {\AA} {394} {625} 

\refitem {Thorstensen, J.R., Ringwald, F.A., Wade, R.A., Schmidt, G.D., 
          Norsworthy, J.E.} {1991} {\AJ} {102} {272} 

\refitem {Wood, M.A., Burke, C.J.} {2007} {\ApJ} {661} {1042}

\refitem {Wood, M.A., Thomas, D.M., Simpson, J.C.} {2009} {\MNRAS} 
         {{\it in press}} {arXiv:0906.2713}

\refitem {Wu, X., Li, Z., Ding, Y., Zhang, Z., Li, Z.} {2002} {\ApJ} {569} {418}

\end {references}

\end{document}